\documentclass[letterpaper,12pt]{article}   
\usepackage{osajnl2} 
\usepackage[draft]{hyperref} 

\begin{document}

\title{Steering between Bloch oscillation and dipole oscillation in parabolic optical waveguide arrays}

\author{Ming Jie Zheng,$^{1,*}$ Yun San Chan,$^1$ and Kin Wah Yu$^{1,2}$}
\address{$^1$Department of Physics, The Chinese University of Hong Kong, Shatin, New Territories, Hong Kong, China}
\address{$^2$Institute of Theoretical Physics, The Chinese University of Hong Kong, Shatin, New Territories, Hong Kong, China}
\address{$^*$Corresponding author: mjzheng@phy.cuhk.edu.hk}

\begin{abstract}
We study the optical oscillations of supermodes in planar optical
waveguide arrays with parabolically graded propagation constant in
individual waveguide interacting through nearest neighbor couplings.
In these arrays, we have identified a transition between a symmetric dipole oscillation (DO) and a symmetry-breaking Bloch oscillation (BO) under appropriate conditions. There exist obvious correspondences between gradon localization and various optical oscillations. By virtue of an analogue between the oscillation of optical system and that of a plane pendulum, we propose a shift of the graded profile to cause a
transition from BO to DO. We confirm the optical transition
by means of Hamiltonian optics, as well as by the field evolution of
the supermodes. The results offer great potential applications in
optical switching, which can be applied to design suitable optical
devices.
\end{abstract}


\maketitle

\newpage

\section{INTRODUCTION}

The propagation and steering of light in optical waveguide arrays
(OWAs) have attracted much interest
\cite{PhysRevLett.103.033902.2009}, since OWAs are good candidates
to realize the optical analogies of electron dynamics
\cite{OptLett.23.1701.1998, PhysRevLett.96.053903.2006,
JOptSocAmB.24.2632.2007, LaserPhoton.Rev.3.243,
PhysRevLett.103.033902.2009}. Among these, Bloch oscillation (BO)
and dipole oscillation (DO) are two important types of optical
oscillations. BO is the oscillatory motion of a particle in a
periodic potential when a finite force is acting on it. Optical BOs
are easier to realize than electronic BOs, since the coherence time
of an optical wave packet is usually much longer than that of an
electronic wave packet \cite{PhysRevLett91.263902.2003}. The optical
equivalent of a finite force can be either a gradient in the
propagation constant \cite{OptLett.23.1701.1998} or a geometric
variation in the structure \cite{PhysRevB.76.195119}. The former has
been achieved by a temperature gradient in thermo-optic polymer
waveguide arrays \cite{PhysRevLett.83.4752.1999}, or a gradient in
the width \cite{PhysRevLett.83.4756.1999} and/or refractive index
\cite{PhysRevLett91.263902.2003}. For the latter many investigations
have been conducted in helical OWAs \cite{PhysRevB.76.195119},
curved OWAs \cite{PhysRevLett.83.963}, chirped photonic crystals
(PCs) \cite{PhysRevB.72.075119.2005}, and other photonic
heterostructures \cite{RevModPhys.78.455}. Tunable photonic BOs have
been realized in nonlinear composite media with a tilted band
structure \cite{OptLett.33.2200}. If the band structure is
parabolic, long-living photonic DOs can be achieved at the bottom of
a parabolic band \cite{OptLett.34.1777}, while at the top of the parabolic band, BOs occur \cite{LaserPhys.16.367.2006}. Since exact BO is defined in linear index gradient, the term BO used in parabolic band is just an approximation. DOs are distinguished from BOs through the different evolution patterns in the wavevector space. If we can combine the advantages of BOs and DOs in the OWAs, the tunability of light propagation will be improved significantly. The key issue is to realize the transition between BO and DO in the optical system. Although both BOs and DOs have been observed under different initial conditions of the cold atoms in
parabolic optical lattices \cite{LaserPhys.16.367.2006}, the transition between them have not been investigated thoroughly. Thus we aim to study the transition between BO and DO in the parabolic optical waveguide arrays (POWAs) in this work.

The parabolic profile of the propagation constant in POWA can be
obtained by the electro-optical effects and careful structural design. We find that there exists a
mechanical analogue between the optical oscillations in POWA and
the mechanical oscillations of a plane pendulum. Optical DO and BO
in POWA are analogous to the libration and rotation of a plane
pendulum, respectively \cite{Goldstein}. The libration and
rotation can be transformed to each other by applying an impulsive
torque, which changes the angular momentum of the pendulum.
Inspiring by this, we propose to shift the center of the parabolic
index profile, which causes a lift of the propagation constant.
Applying this lift-n-shift procedure, the transition between BO
and DO can be studied. The proposed transition between BO and DO
is demonstrated by Hamiltonian optics, and is further confirmed
through the field-evolution analysis, which shows the
propagation of a discrete Gaussian beam along the axis of
waveguides. It is demonstrated that the tunable range of shift distance and phase through the BO-DO transition is wider than that of a
single BO or DO process \cite{PhysRevLett.103.033902.2009,
OptLett.34.1777, OptLett.33.2200}.

\section{MODEL AND FORMALISM}

The POWA consists of $N=100$ waveguides, as shown schematically in Fig.~\ref{fig:POWA}. The array is divided into two zones ($0 \leq z \leq z_1$ and $z_1 < z \leq z_2$) along the longitudinal direction, where there are two parabolic index profiles $H_0$ and $H_1$, respectively. The centers of these two profiles are different so as to realize the BO-DO transition. A feasible experimental realization of the parabolic index profile is proposed based on the previous related experimental work and the improved structural design of OWAs. A linearly varied effective index profile has been realized in AlGaAs waveguide arrays [9]. Similarly, the parabolic index profile can also be obtained by carefully designing the rib width of each waveguide and the spacing between neighboring waveguides.
The gradually varied rib width of individual waveguide corresponds to a graded on-site potential, while the varied spacing between neighboring waveguides may result in a constant or graded coupling constant of the array. The center of the index profile $H_{0}$ when $0 \leq z \leq z_1$ is on the central waveguide, while that of the index profile $H_1$ when $z_1 < z \leq z_2$ is shifted to the right by a certain amount. The shift of profile center from $H_{0}$ to $H_{1}$ can be realized by imposing an additional linear graded profile of propagation constants by using electro-optical effects. The size of each waveguide is in the micrometer scale. However, the real physical parameters should be calibrated through experiments. The plane wave input beam propagates along the axis of the waveguide array, that is, along the $z$ direction. The waveguide array is labeled by $n$ ($n = 1, 2, ..., N$) in the transverse direction.

According to the coupled-mode theory, the evolutionary equation of
modal amplitude $a_n$ in the $n$th waveguide is written as
\begin{equation}\label{eq:modal}
\left[i\frac{d}{dz}+ V_n \right]a_n(z)+a_{n+1}(z)+a_{n-1}(z)=0\,,
\end{equation}
where
$V_n = \alpha [x(n)-S]^2 + \alpha$ is the ``on-site potential'', in which $x(n) = 4[(n-1)/(N-1)-1/2]$ is the rescaled position of the $n$th
waveguide in the transverse direction. The replacement of $x$ instead of $n$ is for convenience in the following calculation. $S$ is the shift of the parabolic index profile relative to $x=0$. While $\alpha = \Delta / C$ and $z=C Z$ are normalized quantities. Here $\Delta$ is the gradient factor of propagation constant, $C$ the coupling constant, and $Z$ the propagation distance of the beam along the axis
of the waveguide. By careful designing the width of each waveguide and spacing between waveguides, $\alpha$ can be kept as a constant. Substituting the solution $a_n^m(z)=u_n^m
\exp({i\beta_m z})$ into Eq.~(\ref{eq:modal}), we have
\begin{equation}\label{eq:amplitude}
\beta_m u_n^m = \left[\alpha (x-S)^2 + \alpha \right] u_n^m + u_{n+1}^m+ u_{n-1}^m\,,
\end{equation}
where $\beta_m$ means the wavenumber of the supermode $m$ and the transverse mode profile is given by a superposition of the mode amplitudes
$u_{n}^{m}$ of the individual waveguides. Equation (\ref{eq:amplitude}) is rewritten in the matrix form
\begin{equation}\label{eq:matrix}
\beta|u\rangle = \textbf{H} |u\rangle\,,
\end{equation}
where the Hamiltonian matrix $\textbf{H}$ is defined as $H_{nn} =
\alpha (x-S)^2 + \alpha$ and $H_{n,n-1}=H_{n,n+1}=1$. The column
vector $|u\rangle$ and $\beta$ denote the eigenvectors and
eigenvalues of $\textbf{H}$, respectively. Using the Hamiltonian
matrix $\textbf{H}$, Eq.~(\ref{eq:modal}) is written as a
$z$-dependent equation
\begin{equation}\label{eq:z-eq}
-i\frac{d}{dz} |u\rangle = \textbf{H} |u\rangle\,.
\end{equation}
It is analogous to the Schr$\ddot{\rm{o}}$dinger equation in quantum system,
\begin{equation}\label{eq:Seq}
-i\frac{d}{dt} |\phi\rangle = \textbf{H} |\phi\rangle\,.
\end{equation}
Here $\hbar$ is taken to be unity. Thus the quantities $(\beta, z)$
in optical waveguide arrays corresponds to $(\omega, t)$ in quantum
system, and we can refer to the functional dependence of $\beta$ on
transverse wavenumber $k$ as the dispersion relation in periodic
optical waveguide arrays. For graded arrays, we can divide the
infinite waveguide arrays into a large number of sub-waveguide
arrays in the transverse direction, each of which can be regarded as
infinite in size and symmetric in translation. Based on the treatment of graded system, we can obtain the band structure approximately as follows. The solution
satisfies the relation $u_{n+1} = u_{n} \exp(i k)$, where $k$ is the
transverse wavenumber. Substituting this relation into
Eq.~(\ref{eq:amplitude}), we obtain the position-dependent
pseudo-dispersion relation
\begin{equation}\label{eq:disp}
\beta(x,k,S) = 2(x-S)^2 + 2(1 + \cos k)\,.
\end{equation}
We have taken $\alpha = 2$ hereafter. Equation (\ref{eq:disp})
resembles the Hamiltonian of a plane pendulum \cite{Goldstein},
\begin{equation}\label{eq:pendulumE}
H(p_{\theta},\theta) = \frac{p_{\theta}^2}{2mL^2} + mgL(1-\cos \theta)\,,
\end{equation}
where $m$ and $L$ are the mass and length of the pendulum,
$p_{\theta}$ and $\theta$ are the angular momentum and angle of
deflection, $g$ is the acceleration due to gravity. Comparing
Eq.~(\ref{eq:disp}) and Eq.~(\ref{eq:pendulumE}), we can see
that $k$ is analogous to $\theta$ while $x$ to the angular
momentum and $\beta$ to the total energy of the system. In this
mechanical analogue, DO corresponds to the libration while BO
to clockwise and anticlockwise rotations about the pivot.

\section{RESULTS}

\subsection{Normal modes and their transitions}

Let us take the original index profile centered at the central waveguide ($S=0$) as an example to analyze the various normal modes and transitions in POWA. By diagonalizing the Hamiltonian matrix $\textbf{H}$, we obtain the eigenvalues and eigenvectors of the system. As described by Eq.~(\ref{eq:disp}), a tilted band is formed between the lower- and upper- limits $\beta(x,k,-\pi)$ and $\beta(x,k,0)$. The normal modes (called gradons) of POWA must be confined between the classical turning points of the tilted band structure. The beating of a few normal modes of nearby eigenvalues gives rise to various
oscillations between the classical turning points. These normal
modes are localized at different positions with different
propagation constants. There is a critical value of the propagation
constant $\beta_{\rm{c}}=\beta(0,0,0)$. Separated by this critical
line, there are three regions on the phase diagram as shown in
Fig.~\ref{fig:PD_MP}(a), which represent three kinds of gradon
modes. At the bottom of the parabolic band, the modes are nondegenerate and localized at the middle part of the array, which are called middle-nondegenerate gradons. At the upper branches of the parabolic bands, the modes are twofold degenerate due to the symmetry. We need to be careful to choose the correct form of the eigenmodes for the twofold degeneracy. In this
sense, we apply a small perturbation to split the twofold
degenerate eigenmodes, and then let the perturbation tend to zero
to obtain the correct form of the eigenmodes. In the right branch, the modes are called right-degenerate gradons, while in the left branch, the modes are called left-degenerate gradons.
The mode patterns of these three gradon modes and a critical mode are shown in the insets of Fig.~\ref{fig:PD_MP}(a). There exist obvious
correspondences between gradon localization and various
oscillations, which is similar to the findings in graded plasmonic
chains~\cite{JApplPhys.106.113307.2009} and graded OWAs \cite{PhysRevA.81.033829}. When $\beta < \beta_{\rm{c}}$, we have DO between two symmetrical classical turning points, which comes from the the contribution of middle-nondegenerate gradons. When $\beta > \beta_{\rm{c}}$, right-degenerate gradons (left-degenerate gradons)
undergo BO on the right (left) hand side of
the array. A transition between DO and BO is possible when $\beta$
is increased beyond $\beta_{\rm{c}}$. To demonstrate the BO-DO
transition clearly, we define the mean position with respect to the
index profile center as follows,
\begin{equation}
\langle x - S \rangle = \langle u_m |x - S| u_m \rangle\,, (m = 1,
2, ..., N)
\end{equation}
where $x-S$ is rescaled position relative to the center of parabolic profile in the transverse direction, and $u_m$ is the $m$th
eigenmode. The mean position
$\langle x - S \rangle$ versus eigenvalue $\beta$ is plotted in
Fig.~\ref{fig:PD_MP}(b). It can be seen that $\langle x - S
\rangle = 0$ for middle-nondegenerate gradons ($\beta < \beta_{\rm{c}}$), $\langle x - S
\rangle < 0$ for left-degenerate gradons ($\beta > \beta_{\rm{c}}$), and $\langle x -
S \rangle > 0$ for right-degenerate gradons ($\beta > \beta_{\rm{c}}$). The rapid
variation of $\langle x - S \rangle$ at $\beta_{\rm{c}} = 4$
indicates the occurrence of BO-DO transition. When $\beta < \beta_{\rm{c}}$, the
single branch of $\langle x - S \rangle$ indicates the symmetric
DOs. When $\beta > \beta_{\rm{c}}$, the two branches of $\langle x
- S \rangle$ indicates the symmetry-breaking BOs. This shows
that $\langle x - S \rangle$ is a viable parameter to show the
BO-DO transition in POWA.

\subsection{BO-DO transition}

We propose a lift-n-shift procedure to shift the parabolic
index profile, which causes the lift of the propagation
constant. Applying this procedure, the transition between BO
and DO can occur. Let us first sketch an example of BO-DO
transition, as shown in Fig.~\ref{fig:BO-DO}(a). In the range
$0 \leq z \leq z_1$, the original index profile is centered at the central waveguide ($S=0$) with Hamiltonian $H_0 = \beta(x,k,0)$. To make the occurrence of BO (DO), the input beam should be a combination of eigenmodes with different propagation constants. In the following explanation, the propagation constants $\beta_{\rm BO}$ and $\beta_{\rm DO}$ in Fig.~\ref{fig:BO-DO} represent the central propagation constants of the beam which undergo BO and DO, respectively. The start point A at $z=0$ has a transverse position $x = \sqrt{2}$ and transverse wavenumber $k = 0$. The dominant modes at point A are
right-degenerate gradons, thus BO occurs at the right side of the array, whose period is $z_{\rm BO}$. After a propagation distance $3/2 z_{\rm BO}$, the beam center reaches a point B, where $z=z_1$, $x=2$, and $k=-\pi$. From the propagation distance $z = z_1$, the index profile is shifted to the right by an amount $S = 1$, that is, $H_1 = \beta(x, k, 1)$. As a consequence, the beam center is moved downward to a point C, where the dominant modes become middle-nondegenerate gradons. Thus DO occurs between points C and
D. Hence, the BO-DO transition is realized by shifting the
center of the index profile. The lift-n-shift procedure can
also be demonstrated in Fig.~\ref{fig:BO-DO}(b). The solid
lines 1, 2, and 3 represent the phase space orbits for DO,
critical motion, and BO respectively for $S = 0$. The dashed
lines 1$^{\prime}$, 2$^{\prime}$, and 3$^{\prime}$ denote the
DO, critical motion, and BO respectively for the shift $S = 1$,
marked by an arrow. The corresponding points of lift-n-shift procedure A $\rightarrow$ B $\rightarrow$ C $\rightarrow$ D are also marked
on Fig.~\ref{fig:BO-DO}(b) accordingly. The mechanical analogue
is useful to analyze the BO-DO transition graphically. The
lift-n-shift procedure is a viable scheme in controlling the
light propagation in optical waveguide arrays. Through the
BO-DO transition, we can realize the optical steering in POWA,
that is, the position of output signal is shifted with respect to
the input signal.

\subsection{Hamiltonian optics}

The proposed switching procedure between BO and DO can be
demonstrated by Hamiltonian optics approach, which is important to the quantum-optical-mechanical analogies. From the position-dependent
dispersion relation Eq.~(\ref{eq:disp}), the evolution of the beam
can be solved by using the equations of motion
\begin{equation}
\frac{dx}{dz} = \frac{\partial \beta(x, k, S)}{\partial k}\,, \qquad
\frac{d k}{dz} = -\frac{\partial \beta(x, k, S)}{\partial x}\,.
\end{equation}
From these equations, the conjugate variables play analogous roles
as action-angle variables in a plane pendulum, where $x$ and $k$
correspond to the angular momentum and angle, respectively. The
equations of motion can be integrated in terms of Jacobi elliptic
functions \cite{Goldstein} and be calculated numerically as well. The numerical Hamiltonian optics results of
the mean transverse position $\langle x \rangle$ and the mean
transverse wavenumber $\langle k \rangle$ are shown by solid lines
in Figs.~\ref{fig:COMP_CP}(a) and \ref{fig:COMP_CP}(b),
respectively. Separated by the line $z = z_1$, there are two zones ($0 \leq z \leq z_1$ and $z_1 < z \leq z_2$) in the whole range of propagation distances. When $0 \leq z \leq z_1$, Fig.~\ref{fig:COMP_CP}(a) shows $\langle x \rangle$
periodically varies on the right side of POWA with increasing
propagation distance $z$, and Fig.~\ref{fig:COMP_CP}(b) shows the
reduced $\langle k \rangle$ in the first Brillouin zone which
indicates that $\langle k \rangle$ increases in the negative
direction with propagation distance $z$. Both features demonstrate that BO occurs in this range. When $z_1 < z \leq z_2$, both $\langle x
\rangle$ and $\langle k \rangle$ varies periodically with the
propagation distance $z$. These features indicate that DO takes
place in this range. Therefore, the results confirm the occurrence of the BO-DO transition, which is realized by shifting the center of index profile. This shift leads to a lift of the propagation constant, that is, the jump of $\beta$ at $z_1$ is caused by the shift of index profile from $S = 0$ to $S=1$ (figure not shown here).

\subsection{Field-evolution analysis}

\begin{figure}[htbp]
  \centering
  \includegraphics[width=0.8 \textwidth]{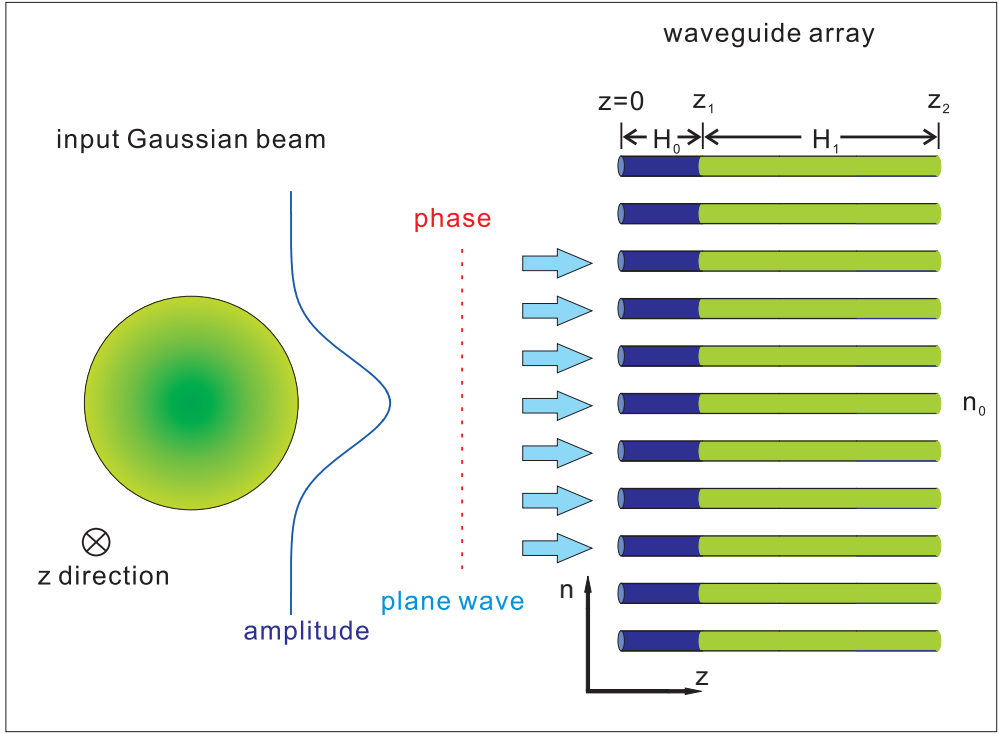}
  \caption{(Color online) Schematic diagram for the parabolic
optical waveguide arrays and the input Gaussian beam. The light
propagates along the axis of waveguide, that is, the $z$
direction. The waveguide array is labeled by $n$ ($n = 1, 2,
..., N$). The parabolic propagation constant is described by
$\beta(x,k,S)$ as Eq.~(\ref{eq:disp}), $H_0 = \beta(x,k,0)$ and
$H_1 = \beta(x,k,1)$ are applied in the corresponding ranges
$[0,z_1]$ and $[z_1, z_2]$, respectively. The input Gaussian
beam has the form as described in Eq.~(\ref{eq:input}), whose cross section is denoted by the green circle. The
parameters are $N = 100$, $n_0 = 86$, $k_0 = 0$, $\sigma = 1$, $z_1 =
1.39$, and $z_2 = 5.32$. POWA.eps.}
  \label{fig:POWA}
  \end{figure}

\begin{figure}[htbp]
  \centering
  \includegraphics[width=0.4 \textwidth]{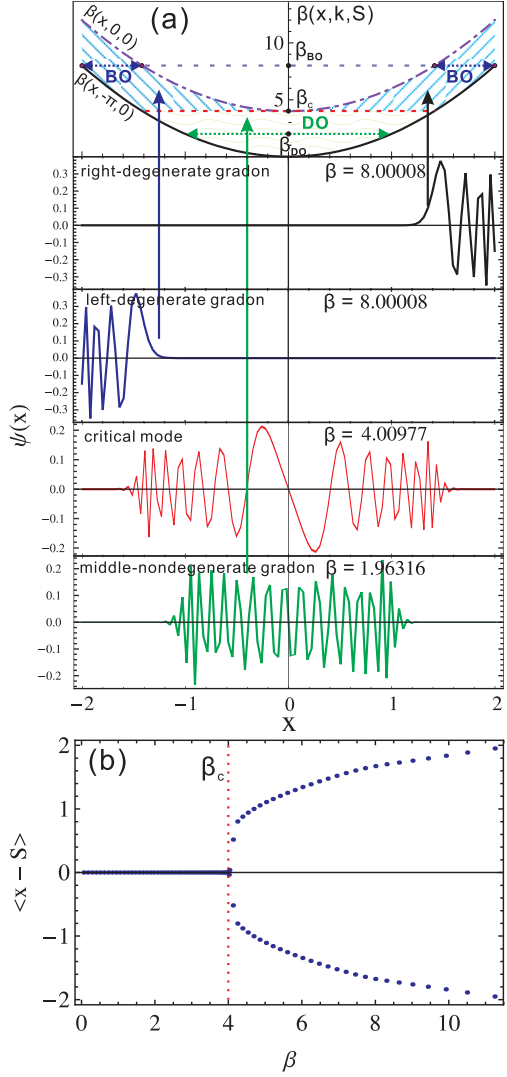}
  \caption{(Color online) (a) Phase diagram for the parabolic optical waveguide arrays with $N=100$ waveguides.
Separated by the critical curve $\beta = \beta_{\rm{c}}$, there
are three regions representing three kinds of gradon modes,
namely the right-degenerate gradons, the left-degenerate gradons, and the middle-nondegenerate gradons. Insets show the mode patterns of the three gradon modes and a critical mode, respectively. (b) The plot of mean position $\langle x - S \rangle$ versus eigenvalues $\beta$ ($S = 0$). The abrupt variation of $\langle x - S \rangle$
indicates the occurrence of BO-DO transition at $\beta_{\rm{c}}
= 4$. PhaseDiagram.eps.}
  \label{fig:PD_MP}
  \end{figure}

\begin{figure}[htbp]
  \centering
  \includegraphics[width=0.6 \textwidth]{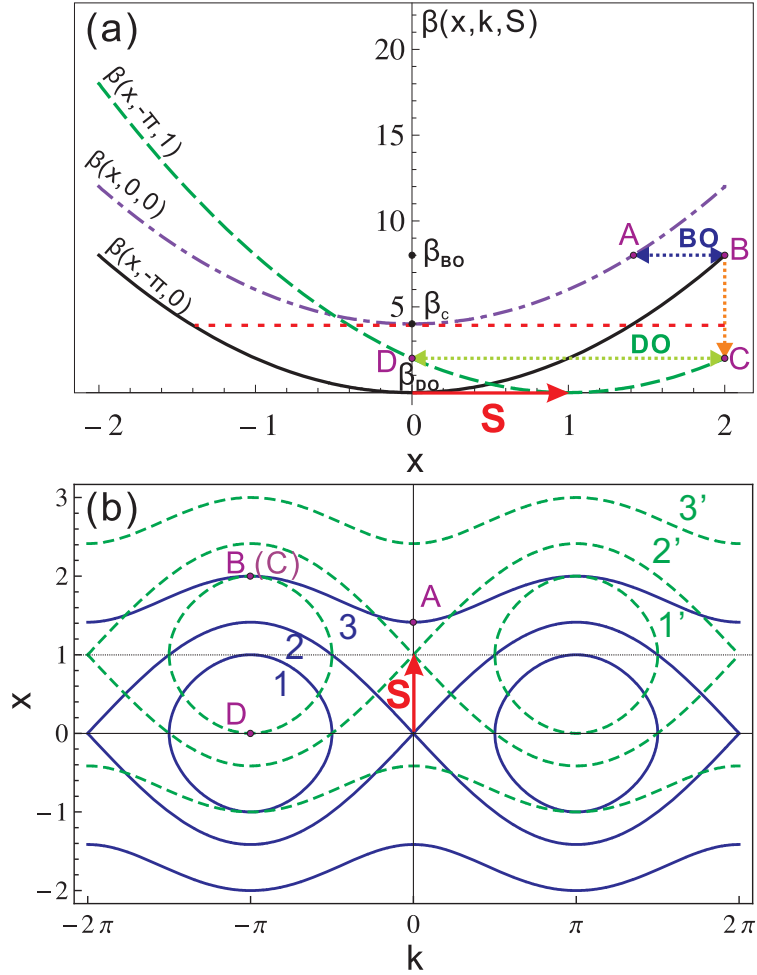}
  \caption{(Color online) (a) A possible BO-DO transition. The arrow marks the shift $S$. The lift-n-shift procedure is shown by the route A $\rightarrow$ B $\rightarrow$ C $\rightarrow$ D. (b) The phase space orbits in POWA for the cases $S = 0$ (solid lines) and $S = 1$ (dashed lines). The solid (dashed) lines 1 (1'), 2 (2'), 3 (3') are corresponding to DO, critical motion, and BO when $S = 0$ ($S = 1$), respectively. The shift $S$ is shown by an arrow. The points A, B, C, D are also marked accordingly. BO-DO.eps.
  }
  \label{fig:BO-DO}
  \end{figure}

The BO-DO transition is further confirmed through the field-evolution analysis. The analysis is performed with an
input wave function at $z = 0$,
\begin{equation}\label{eq:input}
\psi(0)=\frac{1}{(2\pi\sigma^2)^{1/4}} e^{-\frac{(n - n_0)^2}{4
\sigma^2}}e^{-ik_0(n - n_0)}\,,
\end{equation}
where $k_0$ is the input transverse wave number. The incoming
field at $z$ ($z < 0$) is $\psi(z)= \psi(0)\exp(i \beta_0 z)$,
where $\beta_0$ is the propagation constant of individual
homogeneous channel. The intensity profile $|\psi(0)|^2$ has a
discrete Gaussian distribution centered at the $n_0$th
waveguide with spatial width $\sigma$. This input beam is a
discrete Gaussian beam, whose intensity distribution is
schematically shown as the circle in Fig.~\ref{fig:POWA}. The
exponential factor $\exp{[-ik_0(n - n_0)]}$ denotes the phase
differences between input beams excited on the $n$th and the
$n_0$th waveguides. In this study, $k_0 = 0$ represents that
the phase difference between input beams of different
waveguides is zero, that is, the input beam is a plane wave, as
shown in Fig.~\ref{fig:POWA}.

\begin{figure}[htbp]
  \centering
  \includegraphics[width=0.8 \textwidth]{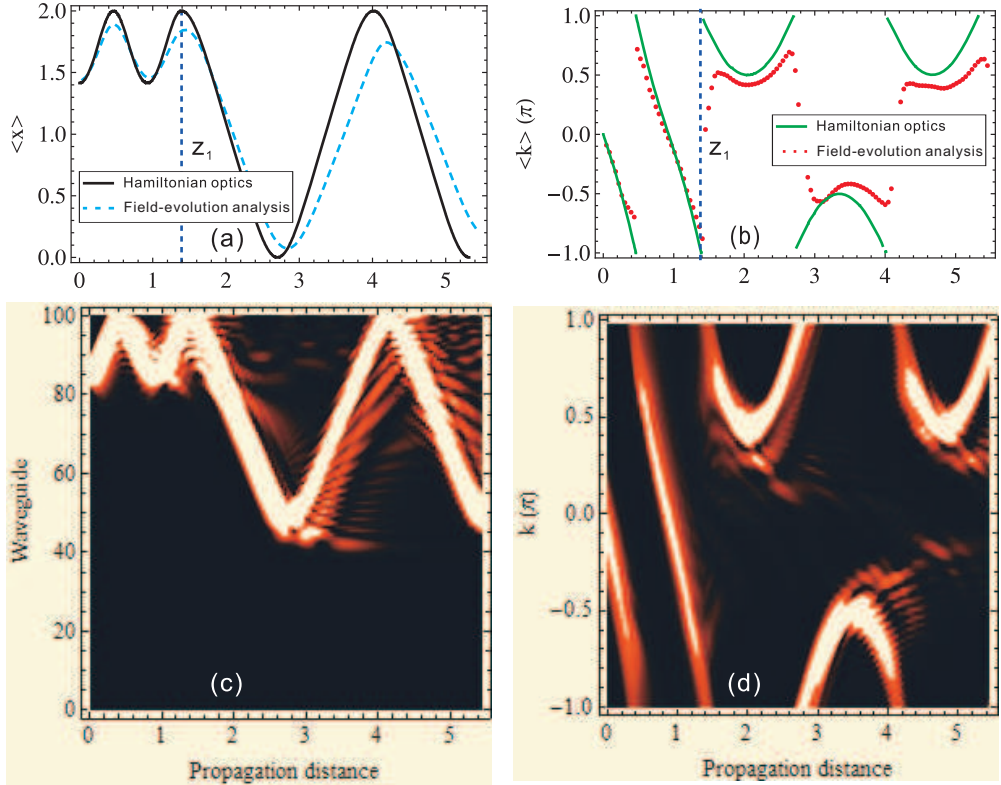}
  \caption{(Color online) Comparison of Hamiltonian
optics results with field-evolution analysis results
for (a) $\langle x \rangle$ and (b) $\langle k \rangle$ in
BO-DO transition. Contour plots of field-evolution analysis results for (c) $|\psi(x)|^2$ as a function of the waveguide index $n$ and the propagation distance $z$ and (d) $|\phi(k)|^2$ as a function of the transverse wave vector $k$ and the propagation distance $z$. xkCP.eps.}
  \label{fig:COMP_CP}
  \end{figure}

We expand the input wave function in terms of supermodes
$|u_m\rangle$,
\begin{equation}\label{eq:expansion}
|\psi (0)\rangle = \sum_{m} A_m |u_m\rangle\,,
\end{equation}
where $A_m = \langle u_m|\psi(0)\rangle$ is the constituent
component of the input Gaussian beam. The subsequent
wave function at propagation distance $z$ is
\begin{equation}\label{eq:beam}
|\psi (z)\rangle = \sum_{m} A_m e^{i\beta_m z}|u_m\rangle\,.
\end{equation}
At a certain propagation distance $z$, the wave function in the reciprocal space can be obtained by taking the following Fourier transform
\begin{equation}\label{eq:phik}
|\phi (k,z)\rangle = \mathcal {F}[|\psi (x,z)\rangle]\,.
\end{equation}
By using these wave functions, the mean value of $x$ and $k$ are
obtained as
\begin{equation}\label{eq:mean}
\langle x \rangle = \frac{\langle \psi|x| \psi \rangle}{\langle
\psi|\psi \rangle} \,, \qquad \langle k \rangle = \frac{\langle
\phi|k|\phi\rangle}{\langle \phi|\phi \rangle} \,.
\end{equation}
Figures \ref{fig:COMP_CP}(a) and \ref{fig:COMP_CP}(b) show the
comparison of Hamiltonian optics results (solid lines) with field-evolution analysis results (dashed
lines) for $\langle x \rangle$ and $\langle k \rangle$,
respectively. At a shorter propagation distance, results of
$\langle x \rangle$ and $\langle k \rangle$ obtained from the two
approaches are in good agreement. As the propagation distance
increases, the discrepancies between them become larger. The
evolution process is clearly demonstrated by the contour plots of
beam intensity $|\psi(z)|^2$ in the real space and $|\phi(k)|^2$
in the reciprocal space, as shown by the contour plots in
Figs.~\ref{fig:COMP_CP}(c) and \ref{fig:COMP_CP}(d),
respectively. Figure \ref{fig:COMP_CP}(c), that is, the contour plot
of $|\psi(z)|^2$ as a function of the waveguide index $n$ and the propagation distance $z$ shows the BO-DO transition in the real space. When the propagation distance increases, there are spurious fields that cannot be lifted and shifted effectively to the required output channels. Since the beam contains many components of modes with different propagation constants, at which the spacing between two classical turning points of the parabolic band are slightly different, the oscillation periods for different modes are different. Another reason is that the force varies slightly in the transverse direction, the different parts of the beam propagate along different waveguides have different oscillation periods. Thus some parts of the beam deviate from the main path after some propagation distances. These are also the reasons for the discrepancies between Hamiltonian optics and field-evolution analysis results. However, the leaked energy in each waveguide is quite small, we can set a threshold for the detection of the output signal, which can avoid the disturbance of the spurious fields.

\section{DISCUSSION AND CONCLUSION}

Our proposed realization of BO-DO transition by lift-n-shift
procedure is advantageous over a single BO or DO process, as
the shift range of BO-DO transition is much larger. To achieve
larger shift of the optical steering, we can combine several steps
of BO-DO or DO-BO transitions. Through the BO-DO transition, we
are able to realize the required position and phase for the
output signal by using appropriate POWA with proper parameters
and boundary conditions.


In summary, we studied the optical oscillations (BO and DO) and
transitions between them in the POWA. The variety of gradon modes and transitions in POWA are identified and the interplay between gradon
localization and various oscillations are elaborated. The mean
position is applied to demonstrate the BO-DO transition in a
set of eigenmodes. We proposed a lift-n-shift procedure to
shift the center of parabolic index profile, which causes a
lift of propagation constant, so that a transition between BO
and DO can occur. The proposed switching procedure between BO
and DO are confirmed by Hamiltonian optics approach and field-evolution analysis. The results from these two methodologies match with each other. Through this kind of switching mechanism, we can
achieve the required position and phase for the output signal
by using appropriate POWA structure with proper parameters and
boundary conditions. These findings have potential applications
in the designing of optical switching devices.

\section*{ACKNOWLEDGMENTS}

This work was supported by RGC General Research Fund of the
Hong Kong SAR Government. We thank Prof. Yakubo for careful
reading of the manuscript and for many useful discussion and
helpful suggestions.



\clearpage

\section*{List of Figure Captions}

Fig. 1. (Color online) Schematic diagram for the parabolic
optical waveguide arrays and the input Gaussian beam. The light
propagates along the axis of waveguide, that is, the $z$
direction. The waveguide array is labeled by $n$ ($n = 1, 2,
..., N$). The parabolic propagation constant is described by
$\beta(x,k,S)$ as Eq.~(\ref{eq:disp}), $H_0 = \beta(x,k,0)$ and
$H_1 = \beta(x,k,1)$ are applied in the corresponding ranges
$[0,z_1]$ and $[z_1, z_2]$, respectively. The input Gaussian
beam has the form as described in Eq.~(\ref{eq:input}), whose cross section is denoted by the green circle. The
parameters are $N = 100$, $n_0 = 86$, $k_0 = 0$, $\sigma = 1$, $z_1 =
1.39$, and $z_2 = 5.32$.

\noindent Fig. 2. (Color online) (a) Phase diagram for the parabolic optical waveguide arrays with $N=100$ waveguides.
Separated by the critical curve $\beta = \beta_{\rm{c}}$, there
are three regions representing three kinds of gradon modes,
namely the right-degenerate gradons, the left-degenerate gradons, and the middle-nondegenerate gradons. Insets show the mode patterns of the three gradon modes and a critical mode, respectively. (b) The plot of mean position $\langle x - S \rangle$ versus eigenvalues $\beta$ ($S = 0$). The abrupt variation of $\langle x - S \rangle$
indicates the occurrence of BO-DO transition at $\beta_{\rm{c}}
= 4$.

\noindent Fig. 3. (Color online) (a) A possible BO-DO transition. The arrow marks the shift $S$. The lift-n-shift procedure is shown by the route A $\rightarrow$ B $\rightarrow$ C $\rightarrow$ D. (b) The phase space orbits in POWA for the cases $S = 0$ (solid lines) and $S = 1$ (dashed lines). The solid (dashed) lines 1 (1'), 2 (2'), 3 (3') are corresponding to DO, critical motion, and BO when $S = 0$ ($S = 1$), respectively. The shift $S$ is shown by an arrow. The points A, B, C, D are also marked accordingly.

\noindent Fig. 4. (Color online) Comparison of Hamiltonian
optics results with field-evolution analysis results
for (a) $\langle x \rangle$ and (b) $\langle k \rangle$ in
BO-DO transition. Contour plots of field-evolution analysis results for (c) $|\psi(x)|^2$ as a function of the waveguide index $n$ and the propagation distance $z$ and (d) $|\phi(k)|^2$ as a function of the transverse wave vector $k$ and the propagation distance $z$.


\clearpage


  \newpage

  \newpage

  \newpage

  \newpage

\end{document}